\documentclass[notitlepage,nofootinbib,
longbibliography,
amsmath,amssymb]{revtex4-1}

\usepackage{graphicx}  
\usepackage{dcolumn}   
\usepackage{bm}        
\usepackage{verbatim}   
\usepackage{bbold}
\usepackage{color}
\usepackage{xcolor}
\usepackage{amsthm}
\usepackage{amsmath}
\usepackage[normalem]{ulem}

\def\D{\Delta}
\def\g{\gamma}
\def\P{\Phi}

\def\e{\epsilon}
\def\be{\begin{equation}}
\def\ee{\end{equation}}
\def\bea{\begin{eqnarray}}
\def\eea{\end{eqnarray}}

\def\>{\rangle}
\def\<{\langle}

\theoremstyle{definition}

\theoremstyle{remark}

\begin{document}

\title{Low temperature $T$-linear resistivity due to umklapp scattering from a critical mode. }

\author{Patrick A. Lee}
\affiliation{
Department of Physics, Massachusetts Institute of Technology, Cambridge, MA, USA
}

\date{December 2, 2020}
\begin{abstract}
We consider the transport properties of a model of fermions scattered by a critical bosonic mode. The mode is overdamped and scattering is mainly in the forward direction. Such a mode appears at the quantum critical point for a electronic nematic phase transition, and in gauge theories for a U(1) spin liquid. It leads to a short fermion life-time, violating Landau's criterion for a Fermi liquid. In spite of this, transport can be described by a Boltzmann equation. We include momentum relaxation by umklapp scattering, supplemented by weak impurity scattering. We find that above a very low temperature which scales with $\Delta_q^3$, where $\Delta_q$ is the minimum umklapp scattering vector, the resistivity is linear in $T$ with a coefficient which is independent of the amount of disorder. We compare the relaxation time approximation with an exact numerical solution of the Boltzmann equation. Surprisingly we find that unlike the resistivity, the Hall Coefficient  strongly deviates from the relaxation time approximation and shows a strong reduction with increasing temperature. We comment on possible comparisons with experiments on high $T_c$ Cuprates.
\end{abstract}


\maketitle

\section{Introduction}
\noindent A great deal of attention has been paid to the linear in temperature resistivity observed in the Cuprate high temperature superconductor family, in the so-called strange metal phase. (For a recent discussion with up to date references, see \cite{legros2019universal}.) A $T$-linear resistivity is not unusual on its own: the standard electron phonon scattering leads to linear in $T$ resistivity at tempertures above approximate 1/4 of the Debye temperature. What is surprising are two addition features. First, the linear $T$ term often extends to low temperatures of a few Kelvin or even less, if the superconductivity is suppressed by a strong magnetic field. There are experiments that show that this low temperature linear term extends over a range of doping, albeit with a doping dependent slope that may or may not match the high temperature slope.\cite{cooper2009anomalous,ayres2020incoherent} A second feature is that at some special doping, the linear $T$ term persists over a large temperature range with the same slope all the way to the lowest temperature. In this paper we shall not address this second feature, but instead focus on the low temperature linear $T$ phenomenon, which is highly unusual on its own, because the standard expectation is $T^2$ behavior.  It is therefore important to ask whether there is a microscopic model that gives rise to the low temperature linear in $T$ resistivity. 
\par

In this paper we consider a model which assumes a two dimensional Fermi sea where the fermions with dispersion $\epsilon_\mathbf{k}$ are being scattered by a critical bosonic mode with the  propagator $P(q,\omega)=1/(Cq^2-i\omega/q)$ so that the spectral function $D(q,\omega)=Im [P(q, \omega)]$ is given by

\begin{equation}
    D(q,\omega)=  \frac{ q \omega}{\omega^2+(Cq^3)^2}  
    \label{Eq: propagator}
\end{equation}
To simplify notations, throughout this paper we measure energy, temperature or frequency in units of the nearest neighbor fermion hopping parameter $t$ and we set the lattice constant to be unity. The fermions are scattered by this mode with a coupling constant $g$.
Equation \ref{Eq: propagator} has arisen in several contexts. For example, this equation describes the critical fluctuations at a  quantum critical point that separates a Fermi surface with four-fold symmetry to one with two-fold symmetry. This is sometimes called the electronic nematic phase transition or the Pomeranchuk instability. This model is considered to be one of the leading candidates for a quantum critical point in the Cuprates. Nematicity is also known to play a prominent role in the iron based superconductors.\cite{hussey2018tale} It is also worth noting that nematicity, $T$-linear resistivity and van Hove singularity are ingredients that are showing up in twisted bilater graphene as well.\cite{cao2020nematicity} Importantly, the singularity is for small momentum $q$. \cite{oganesyan2001quantum,maslov2011resistivity, hartnoll2014transport} This is very different from a charge or spin density wave or an antiferromagnetic instability, where the singularity occurs at a finite $q$.\cite{hlubina1995resistivity,rosch2000magnetotransport} Another example is the scattering of fermionic spinons by U(1) gauge fluctuations that arise in spin-charge separated models in certain quantum spin liquids. \cite{lee1992gauge} This model has  also been discussed in the composite fermion description of the half-filled Landau level. \cite{halperin1993theory} The $\omega/q$ term in the propagator $P(q,\omega)$
 comes from Landau damping of the mode by the Fermi sea. The mode is overdamped with $\omega$ that scales as $q^3$. This leads to a fermion self energy $\Sigma(\omega)$ that goes in leading order as $\omega^{2/3}$, thereby violating the Landau criterion for a well defined quasi-particle. \cite{lee1992gauge} As a result, this model is considered to be an example of a non-Fermi liquid. 

\begin{figure}[htb]
\begin{center}
\includegraphics[width=5in]{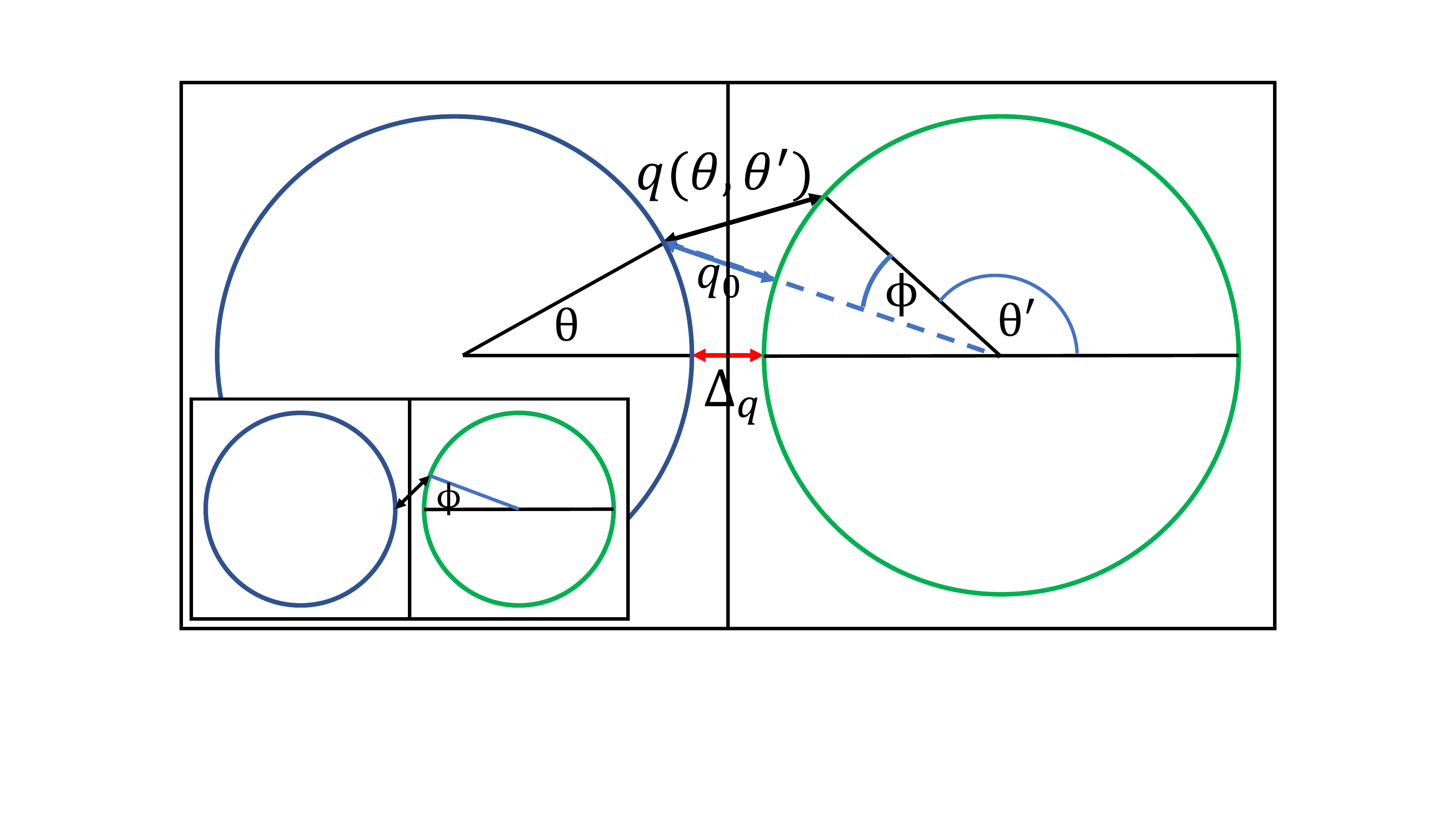}
\caption{Schematic picture of umklapp scattering between an almost circular Fermi surface (colored blue) and its umklapp partner (colored green). This can represent the hole pockets in hole doped Cuprate if the zone center is shifted by $(\pi,\pi)$. A fermion on the Fermi surface parametrized by $\theta$ is scattered to a point parametrized by $\theta'$ on the umklapp Fermi surface by a boson mode carrying momentum $q(\theta.\theta')$. The minimum umklapp vector is $\Delta_q$. $q_0(\theta)$ is the minimum umklapp scattering momentum for the fermion at $\theta$. The picture is repeated for the other three umklapp Fermi surfaces that are closest to the original Fermi surface. The inset shows the special case for the scattering of a fermion located at $\theta=0$ and is useful for the simple calculation of the scattering rate.}
\label{Fig: FS}
\end{center}
\end{figure}

We have earlier suggested that this model gives rise to a transport scattering rate that goes as $T^{4/3}$. \cite{lee1992gauge} This was based on the assumption that the bosonic mode equilibrates to the lattice, but this assumption has been rightly criticized on the grounds that the total momentum is conserved in the scattering process  
and therefore there is no mechanism for relaxing the current. \cite{maslov2011resistivity,hartnoll2014transport} Some form of momentum relaxation such as disorder or umklapp scattering needs to be introduced.  Ref. \cite{maslov2011resistivity}  considered impurity scattering and concluded that the leading T dependent correction for a singly connected convex 2D Fermi surface has a power greater than 2 and saturate for higher temperatures. In ref. \cite{hartnoll2014transport},the problem was worked out for several
versions of disorder  including a linear coupling to the nematic order parameter. In this paper we focus on umklapp scattering as the mechanism of momentum relaxation. The motivation is the following. In the Cuprates, the Fermi level is close to a van Hove singularity so that the minimum umklapp vector $\Delta_q$ connecting neighboring Fermi surfaces is small near the antinode, as seen in fig 1. Due to the $q^3$ scaling in Eq. \ref{Eq: propagator}, the temperature scale when the umklapp scattering kicks in can be very low, being proportional to $\Delta_q ^3$. It will be interesting to find out what the resistivity is above this low temperature scale. A more general remark is that because the mode is overdamped, the umklapp effect is not exponentially small at low $T$ as in the case of electron-phonon interaction, but will in general be power-law down to the lowest temperature. As we shall show, what we find is a $T^2$ behavior that crosses over to $T$ at a very low temperature scale. To see how this come about, we make a quick estimate of the scattering rate $\tilde{\g}$ of a fermion at the umklapp point ( $\theta =0$ ) with energy $T$ by emitting a boson mode with momentum $q$ to reach a final state at $\phi$ on the umklapp Fermi surface, a seen in the inset of Fig. \ref{Fig: FS}
\begin{equation}
\tilde\g \approx \int_{0}^{\infty}d\omega\int d\phi\,(n_0(\omega)+1)  D(q,\omega)   \approx T\int_{0}^{T}d\omega \int d\phi  \,\frac{q}{\omega^2+(Cq^3)^2}
\label{Eq: gamma0}
\end{equation}
In the second step we replaced the Bose factor $n_0(\omega)+1$ by $T/\omega$ and the integration limit by $T$. If $T > C\Delta_q ^3$, we can first perform the $\omega$ integration by extending the upper limit to $\infty$. The integrand is now $1/q^2$. We approximate $q^2 \approx \D_q^2+(k_F\phi)^2$ , and the $\phi$ integration gives a constant. This shows that the scattering rate is proportional to $T/\D_q$. In the opposite case of $T < C\D_q^3$ it is easy to see that the upper limit of the $\omega$ integral gives an additional factor of $T$, leading to a $T^2$ behavior. As mentioned earlier,  unlike the electron-phonon problem, the scattering rate is never exponentially activated.  A more accurate calculation shown later gives the same result. It is worth remarking than since umklapp scattering involves a final state that carries current in the opposite direction, there is no significant distinction between a transport lifetime which is sensitive to the momentum relaxation of each collision and the collision time, in contrast to the case of forward scattering. \cite{lee1992gauge} On the other hand, since the scattering is concentrated at certain points on the Fermi surface, in order to calculate transport properties it is necessary to determine how the momentum distribution relaxes in the vicinity of these points.  We need to solve the Boltzmann equation using the scattering processes as input. This is done in the following sections.

We will calculate the resisitivity, magneto-resisitivty, Hall conductivity and the Hall constant $R_H$. The key physics is that the important scattering process is concentrated near the antinodal point. As such, there is a lot of overlap in the mathematical framework with a different problem involving scattering by antiferromagnetic fluctuations at momentum $(\pi,\pi)$ where strong scattering are concentrated at hot spots. \cite{hlubina1995resistivity} This problem was thoroughly investigated by Rosch \cite{rosch2000magnetotransport} and his work was very helpful to us. In particular he  found a linear T resistivity and linear B magnetoresistivity in certain parameter range. Just as in his case, we found that while some impurity scattering is needed to restrict the modification of the distribution function to the vicinity of a special point on the Fermi surface,  the coefficient of the linear $T$ term is independent of disorder. In this paper we focus on the relatively low temperature regime where the impurity scattering rate is comparable to the inelastic scattering rate, ie. when the temperature dependent part of the resistivity is comparable to the zero temperature limit. We defer discussion of the high temperature limit, or the clean limit, to a later paper.

\section{The Boltzmann equation.}
\noindent  The reader may worry whether one can use a Boltzmann equation when there are no quasi-particles in the Landau sense. The answer is that the Boltzmann equation can be derived without assuming the existence of quasi-particles, as long as the self energy depends on $\omega$ and not on momentum. ( In our problem the leading self energy correction depends only on $\omega$. While a small $k$ dependence was discovered in a higher order expansion \cite{metlitski2010quantum}, we do not expect this to affect the argument in a significant way.) The above- mentioned result  was shown in a classic paper by Prange and Kadanoff.\cite{prange1964transport} Their motivation was that in the electron-phonon scattering problem at temperature high compared with the Deybe temperature, the self-energy is $\pi \lambda \omega$ which can be larger than $\omega$ and yet the Boltzmann equation was used routinely. We have adopted their method to the composite fermion problem \cite{kim1995quantum} and here we sketch the main idea. The key observation by Prange and Kadanoff is that the electron spectral function takes the form

\begin{equation}
    A(\mathbf{k},\omega)=\frac{\Sigma''(\omega)}{(\omega-\epsilon_\mathbf{k} -\Sigma'(\omega))^2+\Sigma''(\omega)^2}
    \label{Eq: spectral}
\end{equation}
where $\Sigma'$ and $\Sigma''$ are the real and imaginary part of the self energy. For $\Sigma''(\omega) \sim$  $\omega^{2/3}$ the spectral function is indeed very broad as a function of $\omega$, but at a given small $\omega$, it is actually a sharp function of $\epsilon_k-\mu$. This distinction is a familiar one in the modern ARPES literature, where the energy dispersion curve (EDC) is broad and the momentum dispersion curve (MDC) is sharp. The conventional derivation of the Boltzmann equation starting from the non-equilibrium Green's function involves integrating over $\omega$. What Prange and Kadanoff proposed to do is to pick a patch of the Fermi surface specified by  $k(\theta)$ which defines a contour along the Fermi surface parametrized by the direction $\theta$ of the Fermi momentum vector and integrate the perpendicular momentum and hence over $\epsilon_k$ instead. As a result, instead of the usual distribution function $f(\mathbf{k})$
they introduced the function $f(k(\theta),\omega)$. Note that in the usual formulation, the vector $\mathbf{k}$ is decomposed into $k(\theta)$ and the momentum in the perpendicular direction, which is then represented by $\epsilon_k-\mu$. Thus we can simply write down the standard Boltzmann equation with the understanding that in the momentum $\mathbf{k}$, the  component perpendicular to the Fermi surface is replaced by $(\epsilon_k-\mu)/v_F$ and $(\epsilon_k-\mu)$ is in turn is replaced by $\omega$. We next introduce in the standard way the linear perturbation from the equilibrium by writing $f(\mathbf{k})$=$f_0(\epsilon_\mathbf{k}) - (df_0(\epsilon_\mathbf{k})/d \epsilon_\mathbf{k})  \tilde\P(\mathbf{k})$= $f^0_\mathbf{k} + f^0_\mathbf{k} (1-f^0_\mathbf{k}) \tilde\P(\mathbf{k})/T $ where  $f_0(\epsilon_\mathbf{k})=f^0_\mathbf{k}$ is the Fermi distribution and $\tilde\P(\mathbf{k})$ has the interpretation of a local shift of the chemical potential at patch $\theta$. This leads to the linearized Boltzmann equation \cite{maslov2011resistivity}:

\begin{equation}
    (e\mathbf{v_\mathbf{k}}.\mathbf{E}+e(\mathbf{v_\mathbf{k}}\times \mathbf{B}). \partial_\mathbf{k} \tilde{\P }) \frac{\partial f_0(\mathbf{\epsilon}_\mathbf{k})}{\partial \mathbf{\epsilon}_\mathbf{k}}= -I_{ee} -I_{ei}
    \label{Eq: BE1}
\end{equation}
where the electron-electron collision term is given by

\begin{align}\label{Eq: collision}\nonumber
I_{ee}=&\sum_{\mathbf{p},\mathbf{q},\mathbf{G}} 2(g^2/T) |P(\mathbf{q}, \epsilon_\mathbf{k}-\epsilon_\mathbf{k-q})|^2[\tilde\P(\mathbf{k})+\tilde\P(\mathbf{p})-\tilde\P(\mathbf{k-q+G})-\tilde\P(\mathbf{p+q})]\\
&\qquad \quad f^0_\mathbf{k}f^0_\mathbf{p}(1-f^0_\mathbf{k-q+G})(1-f^0_\mathbf{p+q}) \delta(\epsilon_\mathbf{k}+\epsilon_\mathbf{p}-\epsilon_\mathbf{k-q+G}-\epsilon_\mathbf{p+q})
\end{align}
where $\mathbf{G}$ denotes a set of reciprocal lattice vectors and the electron-impurity collision term is $I_{ei}=-\gamma_0\tilde\P(\mathbf{k}) (df_0(\epsilon_\mathbf{k})/d \epsilon_\mathbf{k})$ where $\g_0 $ is the impurity scattering rate.  Note that we have described the scattering process as a electron-electron scattering via the exchange of the boson propagator $D(q,\omega)$, rather than the scattering between electrons via the emission and absorption of bosons. The reason is that the conservation of total momentum for the normal scattering ( $\mathbf{G}=0 $) is clearly seen in Eq. \ref{Eq: collision}. The distribution will rapidly equilibrate to a Fermi function with a shifted overall momentum $\tilde{\P}(\mathbf{k}) \propto \mathbf{k.E}$ so that its contribution to the collision integral is exactly zero. Had we considered the emission and absorption of bosons, we would have to keep track of the shifted boson distribution function as well, which introduces more complications. Next, we will show that the vanishing of the normal collision term is more general. As long as the scattering is predominantly forward, then in order to conserve momentum and energy, the scattering partner $\mathbf{p}$ must be approximately equal to either $\mathbf{k}$ or $-\mathbf{k}$. In the first case the last two terms in [ ] in Eq \ref{Eq: collision} cancel the first two terms to first order in $q$. In the second case, since $\tilde{\P}(\mathbf{-k})=-\tilde{\P}(\mathbf{k})$, the first two terms in [ ] in Eq \ref{Eq: collision} approximately cancel. A  similarly argument applies for the next two terms corresponding to the outgoing pair $\mathbf{p+q}$ and $\mathbf{k-q}$. In the presence of impurity scattering which ensures that $\tilde{\P}(\mathbf{k})$ is smoothly varying,  these arguments show that the contribution from normal scattering to the collision integral is negligible. This argument works also in the presence of a magnetic field. The Fermi surface is shifted along an axis that is rotate from the direction of the applied field $\mathbf{E}$ but the relation $\tilde{\P}(\mathbf{-k})=-\tilde{\P}(\mathbf{k})$ continues to hold. So from this point on we will keep only the umklapp term (  $\mathbf{G} \neq 0 $ )
in Eq. \ref{Eq: collision}. On the other hand, in the clean limit, it is necessary to keep the normal scattering because the small residual terms left over from the approximate cancellation control the local equilibration and slow diffusion of the the disturbance of  $\tilde{\P}(\mathbf{k})$ along the Fermi surface towards the umklapp points. This is beyond the treatment of the current paper and is the reason why the results of this paper are limited to relatively low temperatures.

Our next step is to integrate out $\mathbf{p}$ and the radial part of $\mathbf{k}$ taking advantage of the fact that  $df_0(\omega)/d \omega)  $
is a sharply peaked function. To be concrete we choose the origin in $\mathbf{k}$ space so that it is completely enclosed by the Fermi surface and we parametrize the Fermi surface by the angle $\theta$ as shown in Fig \ref{Fig: FS}. Leaving details to the Appendix, we find

\begin{equation}
    e\mathbf{v}_F(\theta).\mathbf{E}+e(\mathbf{v}_F(\theta)\times \mathbf{B}). \frac{\partial \Phi(\theta)}{k_F \partial \theta}= \int \frac{d\theta'}{2\pi} \, 2g^2\nu(\theta')F(\theta,\theta')(\P(\theta)-\P(\theta'))
    \label{Eq: BE2}
\end{equation}
where $\nu(\theta)=k_F/(2\pi v_F)$ is the density of states factor at angle $\theta$. In this equation, all variables are defined on the Fermi surface as parametrized by $\theta$ and  $\P(\theta)$ is defined as $ \tilde{\P}(k(\theta), \omega=0)$.  The kernel of the integral equation $F(\theta,\theta')$ depends on $\theta$ and $\theta'$ only via the magnitude of the momentum transfer $q(\theta,\theta')$ which connects the points $\theta$ and $\theta'$ on the umklapp Fermi surface as shown in Fig. \ref{Fig: FS} . Here we focus on $\theta$ near zero, and the discussion should be repeated for $\theta$ close to the other three nearest neighbor umklapp Fermi surfaces. The kernel is given by

\begin{equation}
F(\theta,\theta')= \int n_0(\omega) (n_0(\omega)+1) (\omega/T) D(q,\omega) \, d\omega
\label{Eq: F}
\end{equation}
where $n_0(\omega)$ is the Bose function.  Note that after integration out the scattering partner, the Boltzmann equation now takes the form of umklapp scattering by emission and absorption of the mode defined by $D(q,\omega).$ The kernel can be written in the form $F(\theta,\theta')=2qf(a)$  where $a = Cq^3/T$ and the scaling function $f(a)$ is given by

\begin{equation}
f(a)= \int_{0}^{\infty} dx \, \frac{e^x} {(e^x-1)^2} \frac{x^2}{x^2+a^2}.
\label{Eq: scaling}
\end{equation}
where $f(a)=\pi/2a$ for small $a$ and $\pi^2/3a^2$ for large $a$. \cite{hlubina1995resistivity} For our numerical work we find it useful to approximate $f(a)$ by the form suggested by Rosch \cite{rosch2000magnetotransport} which matches the large and small $a$ limits and is accurate to 2 \% everywhere:
$f(a)=\pi^2(2\pi+a)/[a(4\pi^2+6\pi a+3a^2)]$.

The coefficient of $\P(\theta$) on the right hand side of  Eq. \ref{Eq: BE2} can be interpreted as the scattering out rate $\gamma(\theta)$ for an electron at $\theta$. 

\begin{equation}
\gamma(\theta)= \int 2 g^2 \nu(\theta') F(\theta,\theta') \, d\theta'/2 \pi
\label{Eq: gamma}
\end{equation}
We now estimate $\gamma(\theta)$ for small $\theta$.
In Fig \ref{Fig: FS} the line connecting $\theta$ to the center of  the next Brillouin zone defines   $q_0(\theta)$ which is the shortest distance from $\theta$ to a point on the umklapp Fermi surface. Then $q(\theta,\theta')$ can be approximated by $\sqrt{q_0^2(\theta)+ (k_F \phi)^2}$ The $\theta'$ integral can be replaced by an integral over $\phi$. For $ T \ll Cq_0^3(\theta)$, we use the form for large $a$ given after Eq.  \ref{Eq: scaling}. The integral over $\phi$ is rapidly convergent and we find $\gamma(\theta) \propto T^2/(v_FC^2 q_0^4(\theta)).$ Note that it is a rapidly decaying function of $q_0(\theta)$.

Next we consider the case $T>Cq_0^3(\theta)$ and use the small $a$ limit of $f(a)$. The $\phi$ integration is easily done to give

\begin{equation}
\gamma(\theta)=\frac{\alpha T}{ q_0(\theta)}   ,\quad     T>Cq_0^3(\theta)
\label{Eq: gammaT}
\end{equation}
where $\alpha=g^2/(2v_F C)$. The linear $T$ dependence of the scattering out rate near the umklapp point which onsets at very low temperature is one of the main results of this paper.

\section{Solution of the Boltzmann equation.}
\noindent

The difficulty of the solution of the Boltzmann equation comes from the second term in Eq \ref{Eq: BE2} which involves an integral over $\P(\theta')$. To make analytic progress, we shall argue that a relaxation time approximation is justified for the resistivity ($B=0$) at low temperatures, which we verify later with a numerically exact solution. Let us consider an electric field $E$ along $x$ and $B=0$. We expect the Fermi surface to be displaced along $x$ and the displacement will satisfy $\P(\theta+\pi)=-\P(\theta)$. From Fig \ref{Fig: FS}, we expect $\theta'$ to be near $\pi$. So we define $\theta''=\theta'-\pi$. Furthermore, since $F$ is peaked near small $q$, we expect $-\theta''$ to be near $\theta$. Since $\P$ is an even function of $\theta$ in the absence of $B$, and we expect it to be slowly varying on the scale of the relatively short range kernel
$F(\theta,\theta')$, we can replace $\P(\theta')$ on the right hand side in Eq \ref{Eq: BE2} by $-\P(\theta)$. This leads to the relaxation time approximation, with a relaxation time of $1/(2\gamma)$. We obtain $\P(\theta)=e\mathbf{v_F.E}/(2\gamma(\theta)+\gamma_0)$ and the current and conductivity can be computed by averaging the product of $\P(\theta)$ and the x component of the Fermi velocity over the Fermi surface in the standard way. For simplicity we will assume a circular Fermi surface with a constant Fermi velocity in the analytic estimate below.

In the absence of impurity scattering, $\P(\theta)$ is strongly suppressed near the umklapp point where the scattering is strong. The current is mainly carried by nodal electrons and current is no longer controlled by the distribution function near the umklapp points. A similar situation was pointed out by Hlubina et al \cite{hlubina1995resistivity} in connection with the hot-spot problem. Just as in that case,\cite{rosch2000magnetotransport} the situation is different once a small amount of impurity scattering is introduced, with scattering rate  $\gamma_0 \ll 1$. The conductivity (resistivity) saturates to a constant $\sigma_0$ ($\rho_0$ ) at zero temperature and we can consider the relative deviation at finite $T$. For small deviation $(\rho(T)-\rho_0)/\rho_0 = -(\sigma(T)-\sigma_0)/\sigma_0  = -\gamma_0 \int  (1/(2 \gamma(\theta)+\gamma_0)-1/\gamma_0 ) \, d\theta /2\pi.$

We will use Eq. \ref{Eq: gammaT}. For $\Delta_q \ll k_F$, $q_0(\theta)$ can be approximated by $\Delta_q + 2 k_F \theta^2. $ The integral over $\theta$ is easily done after extending the integration limits to infinity, a step which is justified as long as $(\rho(T)-\rho_0)/\rho_0<1$.  We obtain

\begin{equation}
    \frac{\rho(T)-\rho_0}{\rho_0} = 
    \frac{\alpha T /(\gamma_0 \sqrt{2k_F}) }{\sqrt{\Delta_q+2 \alpha T/\gamma_0}}.
    \label{Eq: rho}
\end{equation}
This equation should be valid for $ T > C \Delta_q^3$ and when the value of $(\rho(T)-\rho_0)/\rho_0$ is limited to be less than unity. We conclude that the resistivity begins as a constant $\rho_0$ plus a $T^2$ term which quickly changes to a linear $T$ dependence, and then crosses over to a limited region of $\sqrt{T} $ behavior for $T>\Delta_q  \gamma_0 /\alpha$ before the limit set by $(\rho(T)-\rho_0)/\rho_0 < 1$ is reached. Note that the coefficient of the linear $T$ term in the resistivity $\rho(T)$ itself is independent of the impurity scattering rate $\gamma_0$ and is therefore intrinsic. It depends weakly on $\Delta_q$ as $1/\sqrt{\Delta_q}$. At even higher temperatures, the assumption of the conductivity being dominated by small $\theta$ no longer holds and the resistivity increases faster than linear $T$.

\begin{figure}[htb]
\begin{center}
\includegraphics[width=6in]{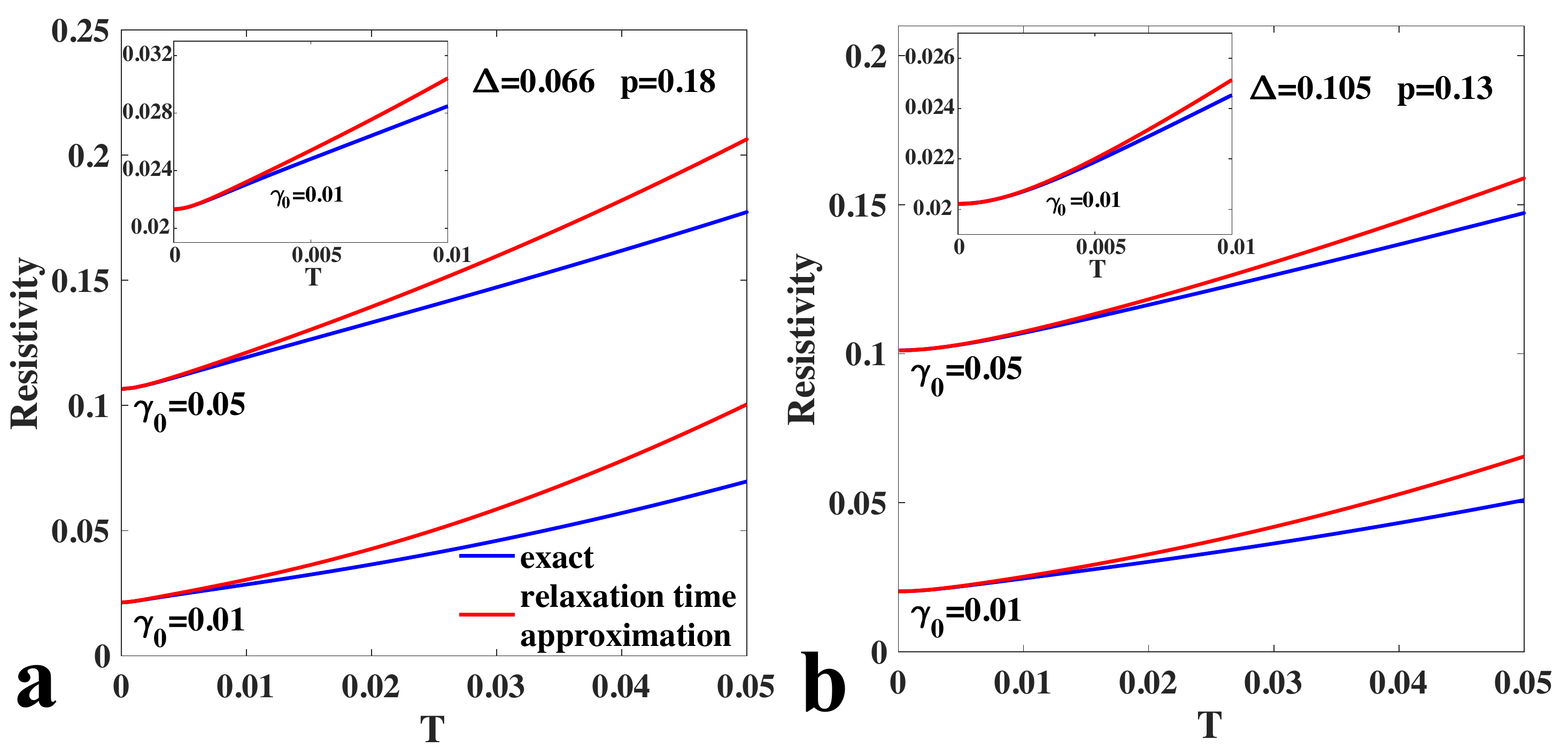}
\caption{The resistivity vs temperature is shown in (a) and (b) for two values of doping $p$. Temperature is measured in units of the nearest neighbor hopping parameter $t$. Note that $\Delta $   is the umklapp vector $\Delta_q $ shown in Fig. \ref{Fig: FS} measured in units of the Brillouin zone dimension, ie.,  $\Delta=\Delta_q/2 \pi$. The impurity scattering rate is $\gamma_0$. The blue lines are obtained by solving the linearized Boltzmann equation, Eq. \ref{Eq: BE2} numerically, while the red lines are obtained in the relaxation time approximation. The inset shows an expanded view of the low temperature region. Note that in (a) an apparently $T$-linear resistivity extends from the lowest temperature up to $T=0.05$ for $\gamma=0.05$. The cross-over from $T$ to $T^2$ can be seen in the inset of (b) which is for a larger $\D$, confirming the expected $\D^3$ dependence of the cross-over scale. }
\label{Fig: resist}
\end{center}
\end{figure}

We next solve the Boltzmann equation numerically without making the relaxation time approximation. This is done by discretizing the $\theta'$ variable in Eq. \ref{Eq: BE2} and converting the integral-differential equation (in the presence of $B$) by a matrix equation. Inverting the matrix numerically gives an accurate solution for $\P(\theta)$. The results for the resistivity are shown in Fig \ref{Fig: resist}. 
We introduce the dimensionless umklapp vector measured in units of the Brillouin zone size $\Delta=\Delta_q/2\pi$. We also re-write $C\Delta_q^3$  as  $\tilde{C}(\Delta_q/k_{F0})^3 $ and in the numerical calculations we set $\tilde{C} = C k_{F0}^3=1$ where $k_{F0}$ is the Fermi wavevector along $\theta=0$. We also set $g=1$ which simply sets the  scale of the temperature dependent part of the  conductivity.
For the band dispersion $\epsilon_\mathbf{k}$ we use the
parametrization suggested by Horio et al. \cite{horio2018three} for Nd doped LSCO, $t'/t= - 0.136, t''/t=0.068$  where $t'$ and $t''$ are the further neighbor
hopping integrals. This model has a van Hove singularity at doping $p \approx 0.22$ beyond which the large hole Fermi surface centered at $(\pi,\pi)$ becomes an electron Fermi surface centered at zero.
Near this doping the dimensionless umklapp vector $\Delta$ is very small. In Fig. \ref{Fig: resist} we see that the relaxation time approximation works very well at low temperature, but lies above the exact results and deviates from linearity sooner as temperature increases. The linear $T$ behavior is very clear at low $T$ from the inset of Fig \ref{Fig: resist} a. For $\Delta=0.066$ the initial $T^2$ 
term is hardly visible, but becomes apparent for the $\Delta=0.105$ case, as expected from the $\D^3$ dependence of the crossover scale. We verify that the resistivity slope is independent of $\gamma_0$ by inspecting Fig \ref{Fig: resist}. The cross-over to $\sqrt{T}$ is rather subtle and can be seen as a  region of convex curvature in the inset of Fig.  \ref{Fig: resist}a. Numerically the linear $T$ like regime extends to about $2 \rho_0$. 

\begin{figure}[htb]
\begin{center}
\includegraphics[width=6in]{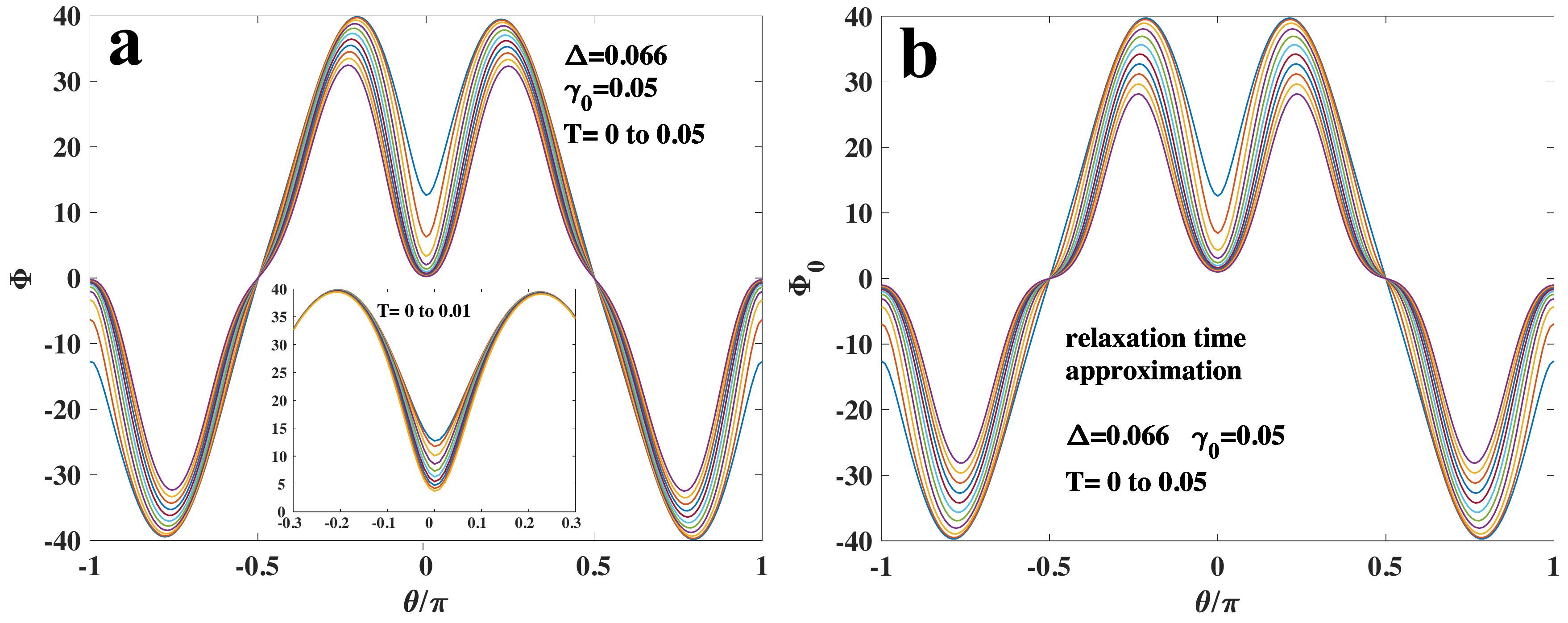}
\caption{(a) The distribution $\P(\theta)$ obtained by solving the Boltzmann equation Eq. \ref{Eq: BE2} numerically for a range of temperatures in ten equally spaced steps. Inset shows the low temperature range in  ten equal steps up to $T=0.01$ (b) same as (a) but using the relaxation time approximation. }
\label{Fig: Phi}
\end{center}
\end{figure}

In Fig \ref{Fig: Phi}a we show $\P(\theta)$ for a range of temperatures for a finite impurity concentration. 
At zero temperature, $\P(\theta)$ is given by the relaxation time approximation due to impurity scattering alone and the dips at $\theta=0$ and $\pi$ are due to the smaller Fermi velocity there.
In the inset, we see that at low temperature, the $\P$ is further suppressed mainly near the umklapp points at $\theta=0$ and $\pi$ so that the current is reduced in a way that is linear in T. As the temperature further increases. the current is increasingly being carried by fermions away from the umklapp points. This is responsible for the eventual deviation from linear $T$  in the resistivity. We also show Fig \ref{Fig: Phi}b the distribution function $\P_0(\theta)$ in the relaxation time approximation. We see that the relaxation time approximation works  very well at low temperature, but tends to over-estimate the effect of temperature as $T$ increases, leading to the upward curvature for the resistivity shown by the red lines in Fig.\ref{Fig: resist}.

\begin{figure}[htb]
\begin{center}
\includegraphics[width=6in]{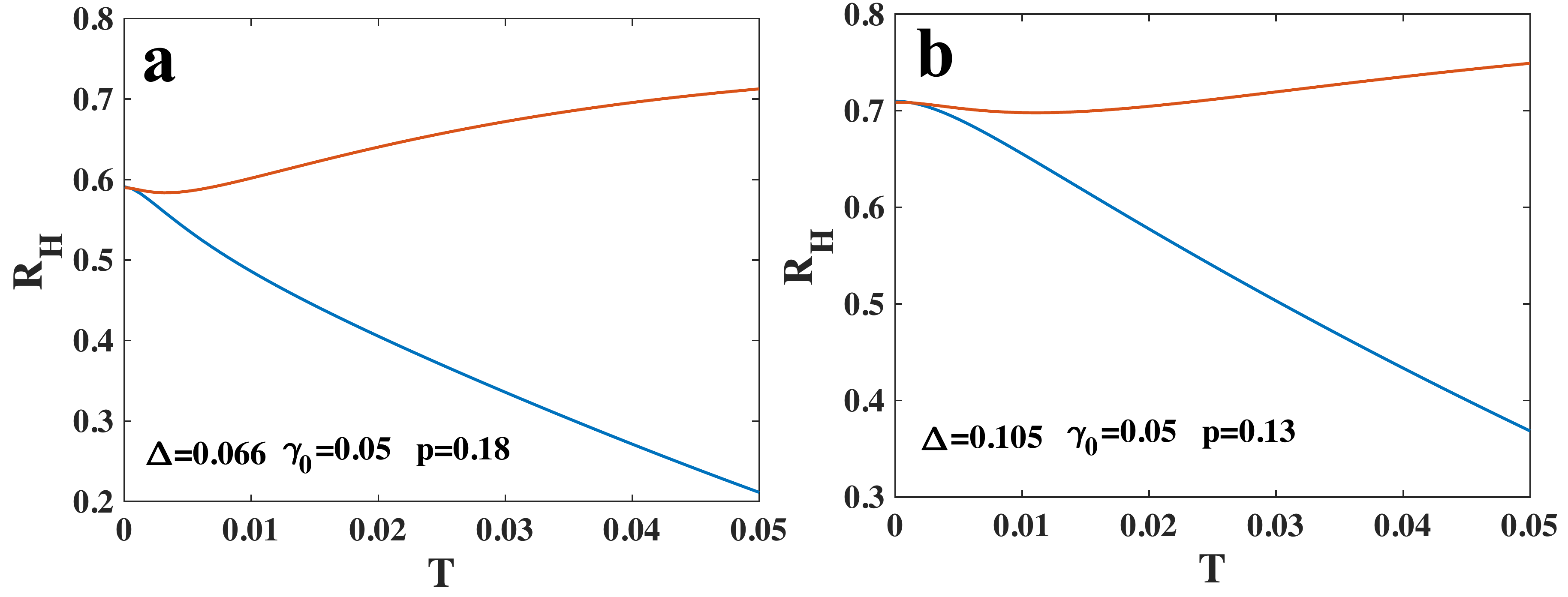}
\caption{The Hall coefficient is plotted as a function of temperature for two values of doping. The blue lines are obtained by solving the Boltzmann equation Eq. \ref{Eq: BE2} numerically while the red line is obtained using the relaxation time approximation. Note that unlike the resistivity, there is a huge discrepancy between the two results. A Hall coefficient that rapidly decreases with increasing temperature is a special feature of our model which is not captured by the commonly used angle dependent relaxation time approximation.}
\label{Fig: RH}
\end{center}
\end{figure}

\begin{figure}[htb]
\begin{center}
\includegraphics[width=5in]{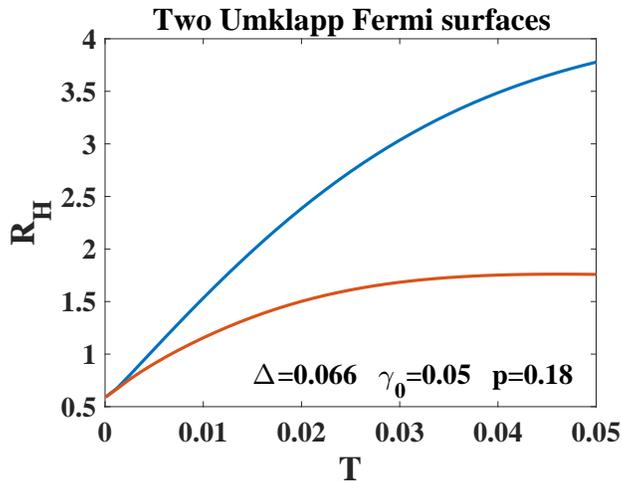}
\caption{Same as Fig. \ref{Fig: RH} a but in a model where there are 2 instead of 4 umklapp Fermi surfaces. For electric field along $x$ the umklapp scattering to the neighboring Fermi surfaces along $y$ has been suppressed. Now $R_H$ increases with increasing temperatures and the trend is qualitatively captured by the relaxation time approximation, shown in red.}
\label{Fig: RH2}
\end{center}
\end{figure}

\begin{figure}[htb]
\begin{center}
\includegraphics[width=6in]{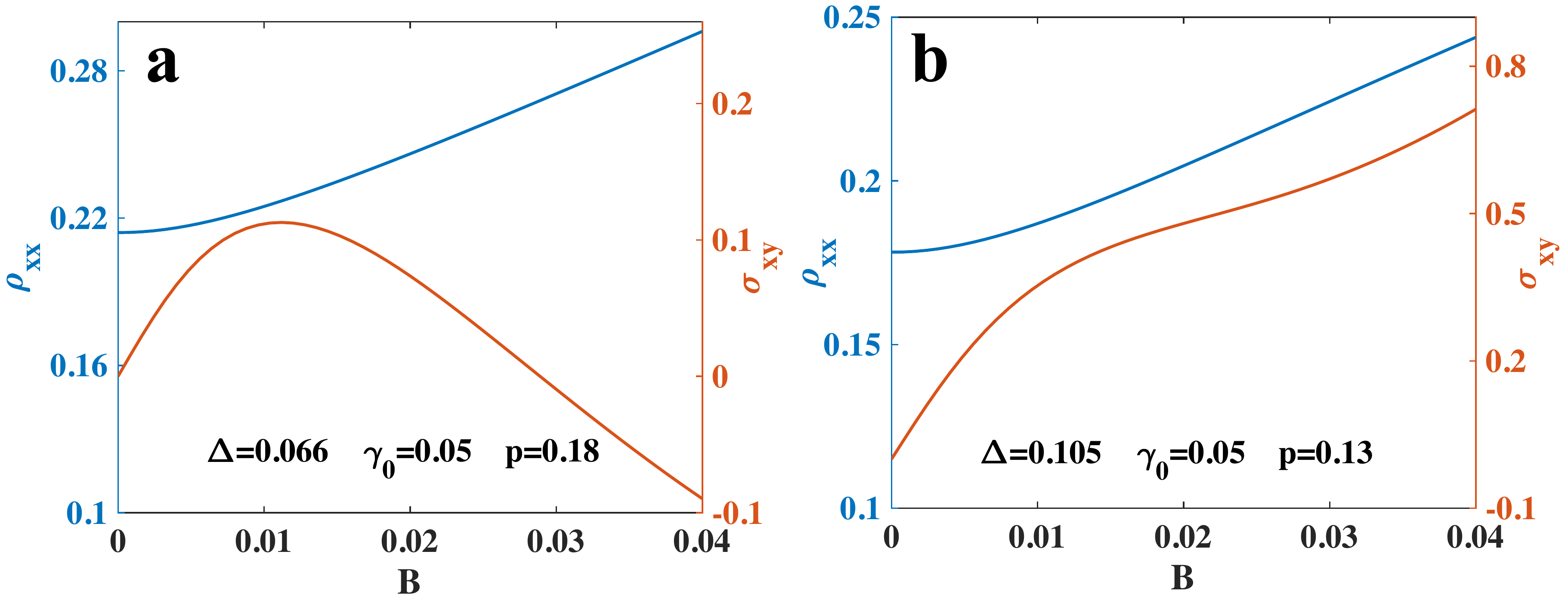}
\caption{The magnetoresistivity $\rho_{xx}(B)$} and the Hall conductivity (in red) for two dopings. While the magnetoresistance looks linear in $B$ above a certain field, that field scale coincides with the deviation of $\sigma_{xy}$ from liearity, which signals the onset of $\omega_c \tau >1 $
\label{Fig: sigma_xy}
\end{center}
\end{figure}

\begin{figure}[htb]
\begin{center}
\includegraphics[width=4in]{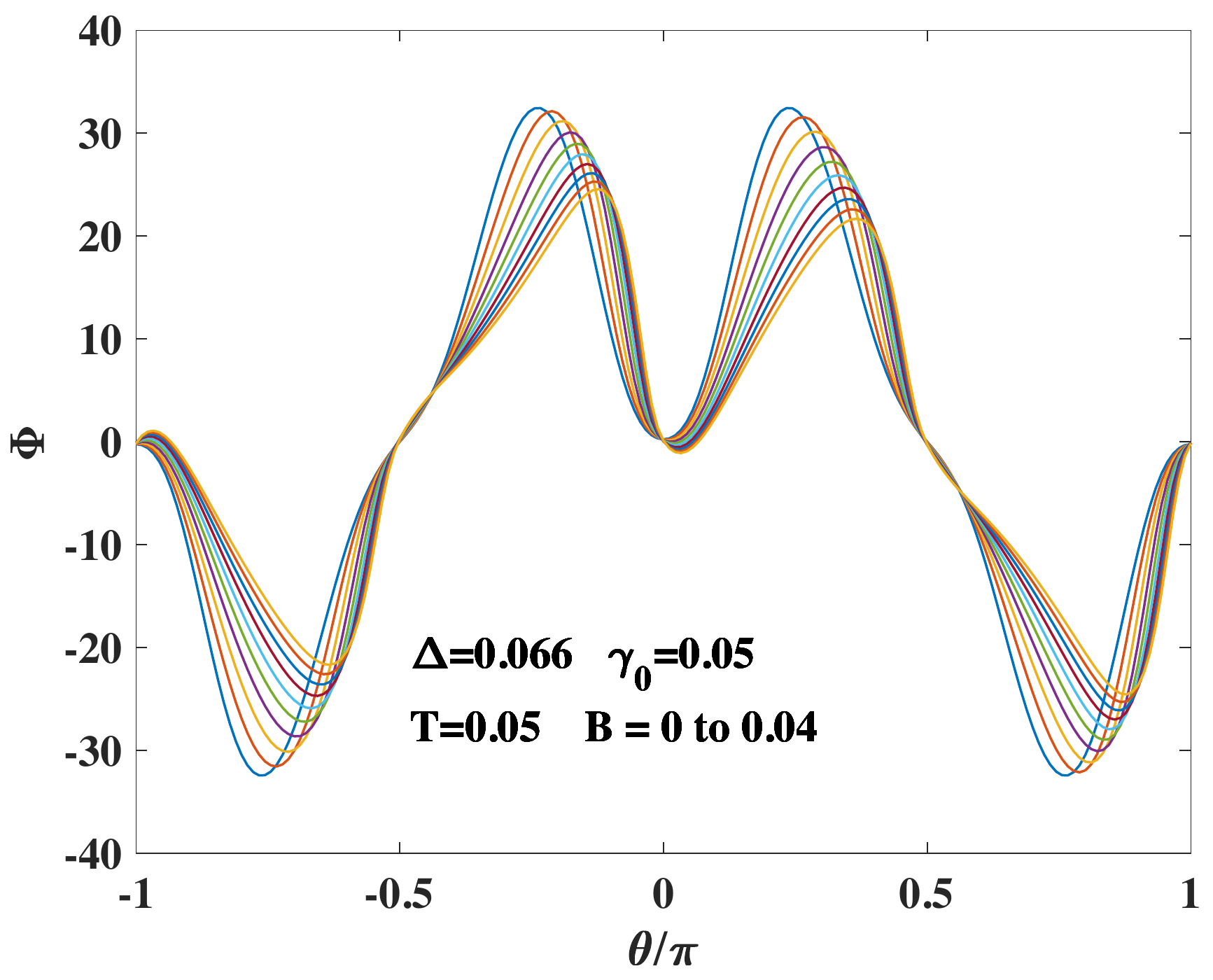}
\caption{The distribution function $\P(\theta)$ for a range of magnetic field B in 10 equal steps.}
\label{Fig: PhiB}
\end{center}
\end{figure}

\section{Magnetic field dependence and possible relations to experiments.}
\noindent

We next discuss the effect of an orbital magnetic field, and compute the Hall effect and magnetoresistance. Our model contains a number of features which have been seen experimentally in a number of Cuprates. Here we show some useful comparisons and also point out a number of inadequacies of our model. 

The Hall coefficient $R_H$ is calculated from the linear B coefficient in the exact solution of the Boltzmann equation and shown by the blue lines in fig. \ref{Fig: RH}. For comparison we also show the relaxation time approximation result, using the standard Zener-Jones formula. \cite{ong1991geometric}

\begin{equation}
\sigma_{xy}= \frac{2e^3B}{h} \int \frac{d\theta}{2\pi}\frac{v_y}{\gamma(\theta)}\frac{d(v_x/\gamma(\theta))}{d\theta}.
\label{Eq: ong}
\end{equation}
The two results agree exactly at zero temperature where impurity scattering dominates, but to our surprise they deviate strongly at higher temperatures. The zero temperature value is a bit lower than the expected $1/(1+p)$. This is because we assume a constant scattering rate as appropriate for Born scattering. If we assume a constant mean-free path as is appropriate for impurities with a fixed scattering cross-section, we do find $1/(1+p)$. This difference is as expected in the presence of anisotropic Fermi velocities.\cite{ong1991geometric} At finite temperatures our model has inelastic scattering  rate which strongly depends on the position on the Fermi surface, being large
near the umklapp point. At a given $T$ its temperature changes from linear $T$ to $T^2$ as $\theta$ moves away from the umklapp point, as seen from Eq. \ref{Eq: gammaT}.  This kind of anisotropic scattering has been used as phenomenological models with considerable success to fit the conductivity and the angle dependent magnetoresistance (ADMR) for the Tl based and other cuprates. \cite{hussey2008phenomenology,putzke2019reduced} They assume an isotropic Fermi liquid like $T^2$ term and a non-Fermi liquid $T$ term that is concentrated near the anti-node. This kind of model bears a lot of similarity to the relaxation time approximation of our more microscopic model. Therefor the resistivity results are similar. Their calculated Hall coefficient also shows a weak temperature dependence, consistent with the red curves in Fig.\ref{Fig: RH}. However, we see that the relaxation time approximation fares poorly for the Hall coefficient, which we find to decrease strongly with increasing temperature. 
As we can see by comparing Fig. \ref{Fig: RH} a and b, the decrease is faster and the discrepancy is stronger for smaller $\D$.
This result suggests the following physical picture. Consider  an electron near $\theta = \pm \pi/2$ in the presence of an  $E$ field along $x$. The umklapp process scatters the electron to a final state on the Fermi surface in the neighboring Brillouin zone in the $y$ direction. This is equivalent to near perfect back scattering and the final state has a y component of the Fermi velocity that is \textit{opposite} in sign to that from a near forward scattering process. This can lead to a reduction and eventually a sign change in the Hall effect as the temperature increases. The information about the final state is encoded in the $
\P(\theta')$ term on the right hand side of Eq. \ref{Eq: BE2} but is absent in the relaxation time approximation. To confirm this picture we have calculated $R_H$ in a model where the umklapp scattering to the neighboring Fermi surfaces located in the $y$ direction have been removed, leaving only two umklapp points along the electric field direction. As seen from Fig. \ref{Fig: RH2} the Hall coefficient is now increasing with $T$ and is qualitatively similar to the relaxation time approximation result. 

Experimentally $R_H$ that decreases with increasing temperatures was observed in underdoped and optimally doped YBCO in the early days of high $T_c$ research and considered to be a major puzzle.\cite{chien1991effect, harris1992experimental} It has never been explained by any theory based on the relaxation time approximation and inspired exotic suggestions such as two different lifetimes for $\sigma_{xx}$ and $\sigma_{xy}$ .\cite{anderson1991hall}  The fact that a strongly temperature dependent $R_H$ shows up due to the breakdown of the relaxation time approximation may give a hint to this long standing puzzle. We should caution that in the experiment $R_H$ is well fitted to a $1/T$ form while our results in Fig. \ref{Fig: RH} is closer to a linearly decreasing function.  Furthermore, this behavior is seen most strongly in the underdoped Cuprates, where $1/R_H \approx p$ . This is a feature of the pseudogap which is not captured in our present model, which gives in the limit of low temperature $1/R_H \approx 1+p$. 

We also computed the Hall conductivity $\sigma_{xy}$ and the magneto-resistivity $\rho_{xx}$ as shown in Fig. \ref{Fig: sigma_xy}. The magneto-resistivity is initially quadratic in $B$ as expected, by switches over to a linear dependence. However, $\sigma_{xy}$ also deviates from linear $B$ at a similar $B$ scale. This indicates a breakdown of the weak field condition $\omega_c \tau < 1$ where $\hbar \omega_c =eB/m$. In a tight binding model we can roughly set $m \approx 1/2t $ so that $B \approx \hbar \omega_c/2t$ which gives us a sense of the size of the dimensionless parameter $B$. For an accessible field of tens of Tesla's, $B$ is indeed of order $10^{-2}$ as shown in Fig. \ref{Fig: RH}. It has been pointed out that a linear $|B|$ magneto-resistance can happen when a group of carriers is moved by the magnetic field to round a corner at $\theta= \pm \pi /2$, where the x component of the Fermi velocity changes sign. This can be a striking effect when the curvature of the turning point is large.\cite{koshelev2013linear} A linear magneto-resistance is also found in a parameter range associated with linear T resistivity by Rosch \cite{rosch2000magnetotransport}. However, it is not known whether $\sigma_{xy}$ also deviates from linearity in these cases.
Experimentally linear $B$ magneto-resistance have been found to be associated with the low temperature linear $T$ behavior in both the Cuprates \cite{giraldo2018scale,ayres2020incoherent} and the iron based superconductors \cite{hayes2016scaling}. It is often (but not always \cite{giraldo2018scale}) fitted to the form $\sqrt{B^2 + T^2}$. However, this is observed in the $\omega_c \tau < 1$ regime and it is not clear if our theory applies. Recent experiments even indicate that the linear in $B$ magneto-resistance may have a Zeeman rather than orbital origin. \cite{ayres2020incoherent}

The effect of a magnetic field on the distribution $\Phi$ is shown in fig. \ref{Fig: PhiB}. We can see the distribution being shifted to the right, but its magnitude also becomes asymmetric as the field is increased. Note that the condition that $\Phi$ is an odd function of $\theta$ is violated by the magnetic field, and the argument given earlier in support of the relaxation time approximation fails.

\section{Conclusion.}
\noindent
We have treated a model of fermions scattered by a critical bosonic mode with umklapp scattering being the dominant momentum relaxation mechanism. After introducing a weak impurity scattering, we find that above a very low temperature scale, the resistivity is linear in $T$. We find that the resistivity is charaterized by a highly anisotropic scattering rate, which is linear in $T$ near the umklapp points and becoming $T^2$ as we move away from the umklapp point. It should be noted this is happening in a model where inverse life-time of the fermion is large, proportional to $T^{2/3}$ and 
Landau's criterion for Fermi liquid is violated. Thus the lifetime of the electron as measured by ARPES is totally different from that inferred from transport measurements. We also find a serious breakdown of the relaxation time approximation in the presence of a magnetic field, leading to a Hall coefficient $R_H$ that decreases with increasing temperature. This is a unique feature (or signature) of transport that is dominated by umklapp scattering.
While many of these features have be seen in Cuprates, we should caution that our model has made a specific assumption that the boson mode is completely soft, i.e. we are either at the quantum critical point or in a critical phase. Deviation from critically will give a low temperature cut-off of the phenomena we describe. We should also remember that the model assumes a sharp fermi surface in the sense that the electron spectral function is sharp in momentum space at low energy, i. e. the MDC as seen in ARPES is well defined. Thus our model does not have  the pseudogap at the antinode, which is a dominant feature in underdoped Cupartes. Recently it was shown that the MDC becomes sharp abruptly for $p$ greater than a certain $p_c$ \cite{chen2019incoherent} and our model may be better suited for this doping range.

{\em Acknowledgment:}
I thank Andrey Chubukov and Dmitrii Maslov for helpful discussions.  This work has been supported by DOE office of Basic Sciences grant number DE-FG02-03ER46076.

\section{Appendix}
\noindent 
 We sketch the derivation of Eq. \ref{Eq: BE2}  starting from Eqs. \ref{Eq: BE1} and \ref{Eq: collision}. The idea is to integrate out the variable $\mathbf{p}$ and replace it with the fermion susceptibility.
 In the case of umklapp scattering, we expect $\mathbf{p}$ to be far from the umklapp points, so that $\tilde{\P}(\mathbf{p})$ is slowly varying and we can drop the term $\tilde{\P}(\mathbf{p})-\tilde{\P}(\mathbf{p+q})$ on the right hand side for small $q$. We also replace $\delta(\epsilon_\mathbf{k}+\epsilon_\mathbf{p}-\epsilon_\mathbf{k-q+G}-\epsilon_\mathbf{p+q})$ with $\int d\omega \, \delta(\epsilon_\mathbf{k}-\epsilon_\mathbf{k-q+G}-\omega)\delta(\omega+\epsilon_\mathbf{p}-\epsilon_\mathbf{p+q})$. We replace the $\mathbf{p}$ integration by an integration along the Fermi surface and one perpendicular to it. For simplicity we shall assume a quadratic dispersion. The radial integration is replaced by an integral over $\epsilon_\mathbf{p} /v_F$. This integration can be done using the identity 
 \begin{equation}
 \int d\epsilon \, f_0(\e-\omega)(1-f_0(\e)) = \omega(1+n_0(\omega)).
\label{Eq: integral}
\end{equation}
The appearance of the Bose factor on the right signals that we are describing a  particle-hole excitation which has Bose statistics. 
Next we do the integration along the Fermi surface which we can replace by $k_F d \theta_p$ where $\theta_p $ specifies a patch on the Fermi surface. For small $q$ we can write $\delta(\omega+\epsilon_\mathbf{p}-\epsilon_\mathbf{p+q}) = \delta (v_\mathbf{p}.\mathbf{q}-\omega)$. We choose the origin of $\theta_p$ to be the point where $\mathbf{q}$ is tangent to the Fermi surface. The delta function then becomes $\delta(v_F q \, sin(\theta_p)-\omega)$. Doing the $\theta_p $ integration gives the factor $\omega/v_F q$ for $\omega < v_F q$, as expected for Landau damping. We obtain
\begin{equation}
 I_{ee} = (g^2/\pi T)\sum_\mathbf{q} \delta(\epsilon_\mathbf{k}-\epsilon_\mathbf{k-q+G}-\omega) (n_0(\omega)+1)f^0_\mathbf{k}(1-f^0_\mathbf{k-q+G})(k_F\omega/v_Fq)|D(q,\omega)|^2[\tilde\P(\mathbf{k})-\tilde\P(\mathbf{k-q+G})]
\label{Eq: Iee}
\end{equation}
Finally we integrate Eq. \ref{Eq: BE1} over $\e_\mathbf{k}$ The lefts side gives the left side of Eq \ref{Eq: BE2}. On the right side  the fermi factors force $\e_\mathbf{k}-\mu$ to be small and we can replace $\tilde{\P}(\mathbf{k})$ by $\P(\theta) $ as defined in the text. The integration over $\e_\mathbf{k}$ is done by using Eq. \ref{Eq: integral} again, and we obtain Eq. \ref{Eq: BE2}.

\bibliographystyle{apsrev4-1}
\bibliography{test}
\end{document}